\documentclass[fleqn,10pt]{wlscirep}

\usepackage[utf8]{inputenc}
\usepackage[T1]{fontenc}

\usepackage{graphicx}%
\usepackage{multirow}%
\usepackage{amsmath,amssymb,amsfonts}%
\usepackage{amsthm}%
\usepackage{mathrsfs}%
\usepackage[title]{appendix}%
\usepackage{xcolor}%
\usepackage{textcomp}%
\usepackage{manyfoot}%
\usepackage{booktabs}%
\usepackage{algorithm}%
\usepackage{algorithmicx}%
\usepackage{algpseudocode}%
\usepackage{listings}%
\usepackage{layouts} 
\usepackage{subcaption}
\usepackage{siunitx}
\usepackage{bm} 			

\title{Shape Optimization of Geometrically Nonlinear Modal Coupling Coefficients: An Application to MEMS Gyroscopes}

\author[1,2,*]{Daniel Schiwietz}
\author[1]{Marian H\"orsting}
\author[2,3]{Eva Maria Weig}
\author[1]{Matthias Wenzel}
\author[1]{Peter Degenfeld-Schonburg}
\affil[1]{Robert Bosch GmbH, Corporate Sector Research and Advance Engineering, Renningen, 71272, Germany}
\affil[2]{Technical University of Munich, School of Computation, Information and Technology, Munich, 80333, Germany}
\affil[3]{Technical University of Munich, Center for Quantum Engineering (ZQE), Garching, 85748, Germany}

\affil[*]{daniel.schiwietz@de.bosch.com}


\begin{abstract}
Micro- and nanoelectromechanical system (MEMS and NEMS) resonators can exhibit rich nonlinear dynamics as they are often operated at large amplitudes with high quality factors and possess a high mode density with a variety of nonlinear modal couplings. Their impact is strongly influenced by internal resonance conditions and by the strength of the modal coupling coefficients. On one hand, strong nonlinear couplings are of academic interest and promise novel device concepts. On the other hand, however, they have the potential to disturb the linear system behavior on which industrial devices such as gyroscopes and micro mirrors are based. In either case, being able to optimize the coupling coefficients by design is certainly beneficial. A main source of nonlinear modal couplings are geometric nonlinearities. In this work, we apply node-based shape optimization to tune the geometrically nonlinear 3-wave coupling coefficients of a MEMS gyroscope. We demonstrate that individual coupling coefficients can be tuned over several orders of magnitude by shape optimization, while satisfying typical constraints on manufacturability and operability of the devices. The optimized designs contain unintuitive geometrical features far away from any solution an experienced human MEMS or NEMS designer could have thought of. Thus, this work demonstrates the power of shape optimization for tailoring the complex nonlinear dynamic properties of MEMS and NEMS resonators.
\end{abstract}
\begin{document}

\flushbottom
\maketitle
\thispagestyle{empty}

\section{Introduction}\label{sec:introductions}
Nonlinear dynamic phenomena are often observed in micro- and nanoelectromechanical system (MEMS and NEMS) resonators. They include simple Duffing-like behavior, nonlinear damping, parametric excitations, or multidmodal effects, just to name a few \cite{bachtold_mesoscopic_2022,eichler_classical_2023,bukhari2021novel,miao2022nonlinearity}. Nonlinear dynamics related to internal resonances between different modes \cite{nayfeh2008nonlinear,kerschen_modal_2014,touze_nonlinear_2006} are particularly relevant in MEMS and NEMS resonators, as shown, e.g., in \cite{opreni_model_2021}. Such effects depend on the strength of nonlinear coupling coefficients and resonance conditions between the modes.

It has been proposed to exploit nonlinearities for novel device concepts. For example, nonlinear modal couplings can be employed for angular rate detection in gyroscopes, enhancement of a gyroscope's sensitivity, frequency stabilization in resonators, energy harvesting, or generation of phononic frequency combs \cite{sarrafan_nonlinear_2019,nitzan_self-induced_2015,antonio_frequency_2012,nabholz_parametric_2020,ganesan_phononic_2017}. In these cases, being able to increase the modal coupling coefficients would enhance favorable effects that arise due to the related nonlinearities.

On the other hand, in typical commercial applications, e.g., gyroscopes or micro mirrors, nonlinear effects related to 3-wave and 4-wave modal couplings can interfere with device functionality and even lead to failure \cite{nabholz_nonlinear_2019,nabholz_spontaneous_2019}. The governing internal resonance conditions are highly sensitive to fabrication tolerances, whereas the involved nonlinear modal coupling coefficients are not \cite{nabholz_validating_2020}. Being able to reduce the modal coupling coefficients sufficiently by design could ensure device functionality despite internal resonances.

In MEMS and NEMS resonators, oscillation amplitudes are often large compared to their structural dimensions. In such a large mechanical deflection scenario, the quadratic displacement components of the mechanical strain become relevant and induce geometric nonlinearities which depend on the mechanical structure and its mode shapes \cite{lifshitz2008nonlinear}. Thus, geometric nonlinearities are one of the main sources of the aforementioned nonlinear phenomena, and structural optimization can be employed to tailor them. So far, the majority of publications on the optimization of MEMS and NEMS resonators was concerned with eigenfrequencies and quality factors \cite{giannini_size_2020,giannini_topology_2020,giannini_topology_2022,shin_spiderweb_2022,hoj_ultra-coherent_2021}. Regarding nonlinear modal couplings, it has been shown that topology optimization can be used to avoid internal resonances in simple structures \cite{pedersen_designing_2005}. Recently, we developed a node-based shape optimization methodology and applied it to a MEMS gyroscope in order to avoid internal resonances \cite{schiwietz_shape_2024}. Alternatively, shape optimization has been applied to tune the geometrically nonlinear Duffing coefficient and a 3-wave coupling coefficient in MEMS resonators \cite{dou_structural_2015}. Based on this approach, designs with optimized Duffing coefficients were fabricated and experimentally verified \cite{li_tailoring_2017}. The approach in \cite{dou_structural_2015} is limited to structures that can be approximated by beam elements and optimizes their width. For example, this implies that an initially straight beam can not become curved. Furthermore, it has been shown that one can also tune nonlinear coefficients of MEMS resonators via topological changes \cite{zega2019hardening}.

In this work, we extend our previously developed node-based shape optimization methodology \cite{schiwietz_shape_2024} to the optimization of geometrically nonlinear 3-wave modal coupling coefficients and exemplify our approach on an industrial MEMS gyroscope. In contrast to our previous work, where we optimized eigenfrequencies, this requires optimization of functions that depend on mode shapes. The advantage of our node-based shape optimization is that it can generate significant shape changes, which go beyond variations in widths. Thus, the presented approach is applicable to arbitrarily complex geometries.

\section{Geometric Nonlinearities}\label{sec:nlgeom}
In this section, we introduce the underlying equations, which govern the dynamics of a mechanical resonator. In particular, we will focus on geometric nonlinearities and the resulting coupling of the structure's eigenmodes.

\subsection{Linear Eigenmodes}
It is useful to formulate the resonator's dynamics in terms of its linear eigenmodes, since they form a basis in which the displacement can be expressed. For that purpose, we employ the finite element method (FEM) to simulate arbitrarily complex geometries. The structure is discretized and the mechanical displacement and geometry are interpolated by shape functions. This leads to the generalized eigenvalue problem
\begin{equation}
\left(\bm{K}-\omega_i^2\bm{M}\right)\bm{\phi}_i=\bm{0}, \label{eq:gevp}
\end{equation}
with linear stiffness matrix $\bm{K}$, mass matrix $\bm{M}$, eigenvector $\bm{\phi}_i$, angular eigenfrequency $\omega_i=2\pi f_i$ and eigenfrequency $f_i$ of mode $i$. A modal analysis is performed to obtain the relevant eigenmodes from Eq.~\eqref{eq:gevp}. The eigenvectors are then mass-normalized, i.e., $\bm{\phi}_i^T\bm{M}\bm{\phi}_i=1$.

\subsection{Geometrically Nonlinear Dynamics}
MEMS and NEMS resonators are typically driven to oscillate at relatively large amplitudes such that geometrically nonlinear effects are not negligible. These effects are solely determined by the geometry of the resonator and its material parameters. We assume linear material behavior, which is customary for standard MEMS and NEMS materials, such as, e.g., silicon. Starting from continuum mechanics, the strain energy $U$ of an elastic body is given by
\begin{equation}
U=\frac{1}{2}\int_V\bm{S}:\bm{E}\,\mathrm{d}V, \label{eq:strainenergy}
\end{equation}
with the second Piola-Kirchhoff stress tensor $\bm{S}$, the Green-Lagrange strain tensor $\bm{E}$ and the undeformed volume of the body $V$ \cite{wriggers_nonlinear_2008}. Note that Eq.~\eqref{eq:strainenergy} contains terms that are quadratic, cubic and quartic in the displacement $\bm{u}$. The displacement can be written in the basis of the linear eigenmodes as
\begin{equation}
\bm{u}=\sum_{i=1}^{N}q_i\bm{\phi}_i, \label{eq:modalsuperposition}
\end{equation}
where $N$ is the number of degrees of freedom in the discretized model and $q_i$ is the modal amplitude of mode $i$. Inserting the modal superposition, Eq.~\eqref{eq:modalsuperposition}, into the strain energy, Eq.~\eqref{eq:strainenergy}, one obtains
\begin{equation}
U=\sum_{n=1}^{N}\frac{1}{2}\omega_n^2q_n^2+\sum_{n,m,l=1}^{N}\alpha_{n,m,l}q_nq_mq_l+\sum_{n,m,l,k=1}^{N}\beta_{n,m,l,k}q_nq_mq_lq_k, \label{eq:strainenergy_modal}
\end{equation}
where $\alpha_{n,m,l}$ and $\beta_{n,m,l,k}$ are modal 3-wave and 4-wave coupling coefficients that arise due to geometric nonlinearities. This work focuses on the 3-wave coupling coefficients $\alpha_{n,m,l}$. Within the FEM approximation they can be calculated as 
\begin{equation} \label{eq:alpha}
\alpha_{n,m,l}=\sum_e\int_{V^e}\left(\bm{\varepsilon}^e_n\right)^T\bm{D}\bm{\eta}_{m,l}^e\,\mathrm{d}V,
\end{equation}
where $e$ is the element index and the summation is performed over all elements, $V^e$ is the undeformed volume of element $e$, $\bm{\varepsilon}^e_n$ is the linear strain of eigenvector $n$ within element $e$, $\bm{D}$ is the elasticity matrix and $\bm{\eta}_{m,l}^e$ is the nonlinear strain of modes $m$ and $l$ within element $e$. Note that Voigt notation is used here by collecting the components of the symmetric strain tensors in vectors $\bm{\varepsilon}^e_n$ and $\bm{\eta}_{m,l}^e$. The linear and nonlinear strains can be obtained as
\begin{align}
\bm{\varepsilon}^e_n&=\bm{B}^e_{\varepsilon}\bm{\phi}^e_n, \\
\bm{\eta}^e_{m,l}&=\frac{1}{2}\bm{B}^e_{\eta}\left(\bm{\phi}^e_m\right)\bm{\phi}^e_l,
\end{align}
where $\bm{B}^e_{\varepsilon}$ contains shape function derivatives and is identical to the strain-displacement matrix of the linear theory and $\bm{B}^e_{\eta}\left(\bm{\phi}^e_m\right)$ contains products of shape function derivatives and derivatives of $\bm{\phi}^e_m$, which is the eigenvector of mode $m$ inside element $e$. The structures of $\bm{B}^e_{\varepsilon}$ and $\bm{B}^e_{\eta}$ can be found in literature \cite{wriggers_nonlinear_2008}.

An equation of motion can be derived from Eq.~\eqref{eq:strainenergy_modal} for each mode, which determines its time-dependent modal amplitude and reads
\begin{equation}
\ddot{q}_n+\frac{\omega_n}{Q_n}\dot{q}_n+\omega_n^2q_n+\sum_{m,l=1}^{N}\tilde{\alpha}_{n,m,l}q_mq_l+\sum_{m,l,k=1}^{N}\tilde{\beta}_{n,m,l,k}q_mq_lq_k=F_n, \label{eq:eom}
\end{equation}
where $Q_n$ is the quality factor of mode $n$, $F_n$ is the modal force acting on mode $n$ and we defined
\begin{align}
\tilde{\alpha}_{n,m,l}&=\alpha_{n,m,l}+\alpha_{l,n,m}+\alpha_{m,l,n}, \\
\tilde{\beta}_{n,m,l,k}&=\beta_{n,m,l,k}+\beta_{k,n,m,l}+\beta_{l,k,n,m}+\beta_{m,l,k,n}.
\end{align}
Equation~\eqref{eq:eom} contains quadratic and cubic nonlinearities, which couple mode $n$ to all other modes. The significance of these couplings is determined by resonance conditions between the eigenfrequencies of the modes and the magnitude of the corresponding coupling coefficients. A detailed description of many of the nonlinear effects that can arise from Eq.~\eqref{eq:eom} can be found, e.g., in \cite{nayfeh2008nonlinear}. To facilitate the numerical modelling of the underlying systems with potentially large FEM models, efficient approaches have been developed to simulate the dynamics arising due to Eq.~\eqref{eq:eom} in a reduced order model (ROM) \cite{touze_model_2021}.

\section{Shape Optimization of Modal Coupling Coefficients}\label{sec:optimization}
In this section, we give an overview of the employed shape optimization methodology. Particular attention is being paid to the sensitivity analysis of eigenvector-dependent functions, such as the 3-wave coupling coefficients. For that purpose, the adjoint method is introduced.

\subsection{Shape Optimization}
We build upon the shape optimization methodology that we introduced previously \cite{schiwietz_shape_2024}. A detailed explanation of the approach can be found therein and we will only give a brief summary here.

We apply node-based shape optimization to geometries that are extruded along the z-direction. For that purpose, parameters $p_j$ are introduced, which shift the exterior nodes of the geometry along the surface normals in the xy-plane. The design parameters are collected in a vector $\bm{p}$. In order to avoid sharp kinks in the optimized design, the parameters are defined such that they also drag neighboring nodes along in a linearly decaying manner. Essentially, each parameter $p_j$ corresponds to one exterior node of the geometry. Increasing the value of $p_j$ will shift the corresponding node, as well as some neighboring nodes, along the geometry's surface normal in the xy-plane. By that, the shape of the extruded geometry can be varied in a parametrized way and gradients with respect to such shape changes can be calculated analytically.

We are concerned with optimization problems where the objective function and the constraints depend on parameters explicitly as well as implicitly through eigenfrequencies and eigenvectors. These optimization problems are generally nonlinear and require an iterative solution procedure. The aim is to find the parameter set $\bm{p}$, which modify the shape of the geometry such that all specified constraints are fulfilled. Therefore, our optimization loop starts by performing a modal analysis of the current design. Subsequently, the objective function and constraints are evaluated and their sensitivities with respect to the design parameters are calculated. The method of moving asymptotes (MMA) \cite{svanberg_method_1987} is then employed to solve the nonlinear optimization problem and provide an updated set of design parameters, based on the function values and sensitivities. The design parameters are then applied to shift the positions of the node coordinates. By that, an updated design is obtained. The steps are repeated until all constraints are satisfied.

\subsection{Adjoint Method}
We wish to optimize functions
\begin{equation} \label{eq:generalfunction}
c=c(p_{j},f_{i}(p_j),\bm{\phi}_i(p_j)),
\end{equation}
which have explicit and implicit dependencies on a given parameter $p_j$. In general, a function $c$ can depend on any number of parameters $p_{j}$, eigenfrequencies $f_{i}$ and eigenvectors $\bm{\phi}_i$. Evaluating the sensitivity $\frac{\mathrm{d}c}{\mathrm{d}p_{j}}$ by calculating the eigenvector sensitivity $\frac{\mathrm{d}\bm{\phi}_i}{\mathrm{d}p_{j}}$, for potentially thousands of parameters, is inefficient. The adjoint method can be applied to evaluate $\frac{\mathrm{d}c}{\mathrm{d}p_{j}}$, without having to obtain the eigenvector sensitivity, leading to a reduced computational complexity \cite{lee_adjoint_1999}. For that purpose, the Lagrangian $\mathcal{L}$ is defined by adding zeros to Eq.~\eqref{eq:generalfunction} as
\begin{equation}
\mathcal{L}=c(p_{j},f_{i}(p_j),\bm{\phi}_i(p_j))+\sum_{i}\left[\bm{\lambda}_{i}^T\left(\bm{K}-\omega_i^2\bm{M}\right)\bm{\phi}_i +\frac{1}{2}\eta_i\left(\bm{\phi}_i^T\bm{M}\bm{\phi}_i-1\right)\right], \label{eq:lagrangian}
\end{equation}
where $\bm{\lambda}_{i}$ and $\eta_i$ are the adjoint variables and the index $i$ runs over all modes on which $c$ depends. The adjoint variables can be chosen arbitrarily, since they are being multiplied by terms which equal zero. Therefore, Eqs.~\eqref{eq:generalfunction} and \eqref{eq:lagrangian} have identical values. However, calculating the sensitivity of Eq.~\eqref{eq:lagrangian} is more efficient and can be obtained as
\begin{equation}
\frac{\mathrm{d}\mathcal{L}}{\mathrm{d}p_{j}}=\frac{\partial c}{\partial p_j}+\sum_{i}\left[\bm{\lambda}_{i}^T\left(\frac{\partial\bm{K}}{\partial p_j}-\omega_i^2\frac{\partial\bm{M}}{\partial p_j}\right)\bm{\phi}_i+\frac{1}{2}\eta_i\bm{\phi}_i^T\frac{\partial\bm{M}}{\partial p_j}\bm{\phi}_i\right], \label{eq:lagrangiangradient}
\end{equation}
where the adjoint variable $\eta_i$ has to be calculated according to
\begin{equation}
\eta_i=-\bm{\phi}_i^T\frac{\partial c}{\partial \bm{\phi}_i},
\end{equation}
and the adjoint variable $\bm{\lambda}_{i}$ has to be obtained by solving the equation system
\begin{align}
\left(\bm{K}-\omega_i^2\bm{M}\right)\bm{\lambda}_i&=-\frac{\partial c}{\partial \bm{\phi}_i}-\eta_i\bm{M}\bm{\phi}_i, \label{eq:nelson1} \\
\bm{\phi}_i^T\bm{M}\bm{\lambda}_i&=\frac{1}{4\pi\omega_i}\frac{\partial c}{\partial f_i}. \label{eq:nelson2}
\end{align}
Equations~\eqref{eq:lagrangiangradient}-\eqref{eq:nelson2} were obtained by taking the derivative of Eq.~\eqref{eq:lagrangian} and then choosing the adjoint variables such that all terms proportional to $\frac{\mathrm{d}\bm{\phi}_i}{\mathrm{d}p_{j}}$ and $\frac{\mathrm{d}f_i}{\mathrm{d}p_{j}}$ cancel out. We solve Eq.~\eqref{eq:nelson1} and Eq.~\eqref{eq:nelson2} with Nelson's method \cite{tcherniak_topology_2002,nelson_simplified_1976}. One can also solve an equation system for $\eta_i$ and $\bm{\lambda}_{i}$ simultaneously \cite{lee_adjoint_1999,dou_structural_2015}, but this would lead to a coeffcient matrix with an increased bandwidth and is less efficient. For all functions that depend on eigenvectors we will use the adjoint method according to Eqs.~\eqref{eq:lagrangiangradient}-\eqref{eq:nelson2} for sensitivity analysis. The required partial derivatives $\frac{\partial c}{\partial p_j}$, $\frac{\partial c}{\partial \bm{\phi}_i}$ and $\frac{\partial c}{\partial f_i}$ are derived in Appendix~\ref{sec:A_derivatives_1} for the 3-wave coupling coefficients. Details regarding the calculation of the stiffness and mass matrix sensitivities, which appear in Eq.~\eqref{eq:lagrangiangradient}, can be found in \cite{schiwietz_shape_2024}.

\section{Application to a MEMS Gyroscope}
In this section, we briefly describe the MEMS gyroscope, on which we will demonstrate our methodology. Furthermore, we formulate two optimization problems, to demonstrate the capability of tuning geometrically nonlinear 3-wave coupling coefficients by shape optimization. We choose an exemplary MEMS gyroscope, to highlight that, in contrast to previous work \cite{dou_structural_2015}, our method requires no simplifications of the model, such as using beam-theory, and is applicable to MEMS structures with a realistic complexity. The application to MEMS gyroscopes holds also practical relevance, as the occurence of nonlinear modal couplings can be detrimental to their operation \cite{nabholz_nonlinear_2019,nabholz_spontaneous_2019}.

\subsection{MEMS Gyroscope Model}
\begin{figure*}
	\centering
	\includegraphics[width=0.9\textwidth]{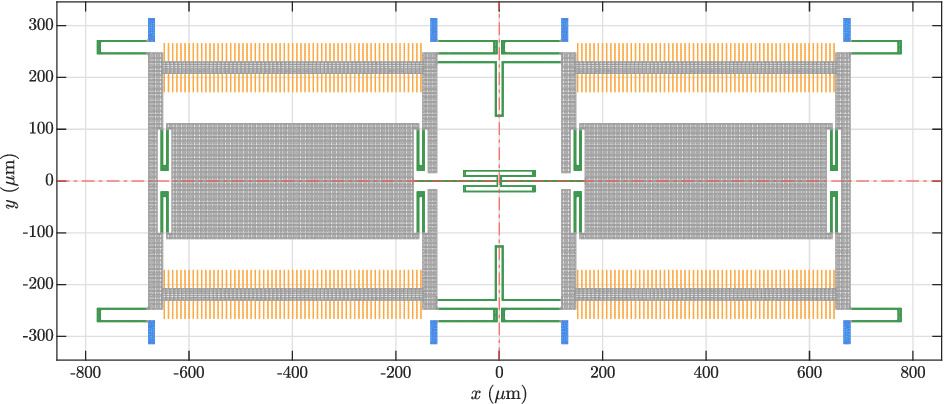}
	\caption{Top view of the single-axis MEMS gyroscope's initial design. The dashed red lines show the symmetry axes, blue elements indicate the substrate anchors, green elements the springs and orange elements the comb electrodes.}
	\label{fig:InitialDesign}
\end{figure*}
\begin{figure*}
	\centering
	\includegraphics[width=\textwidth]{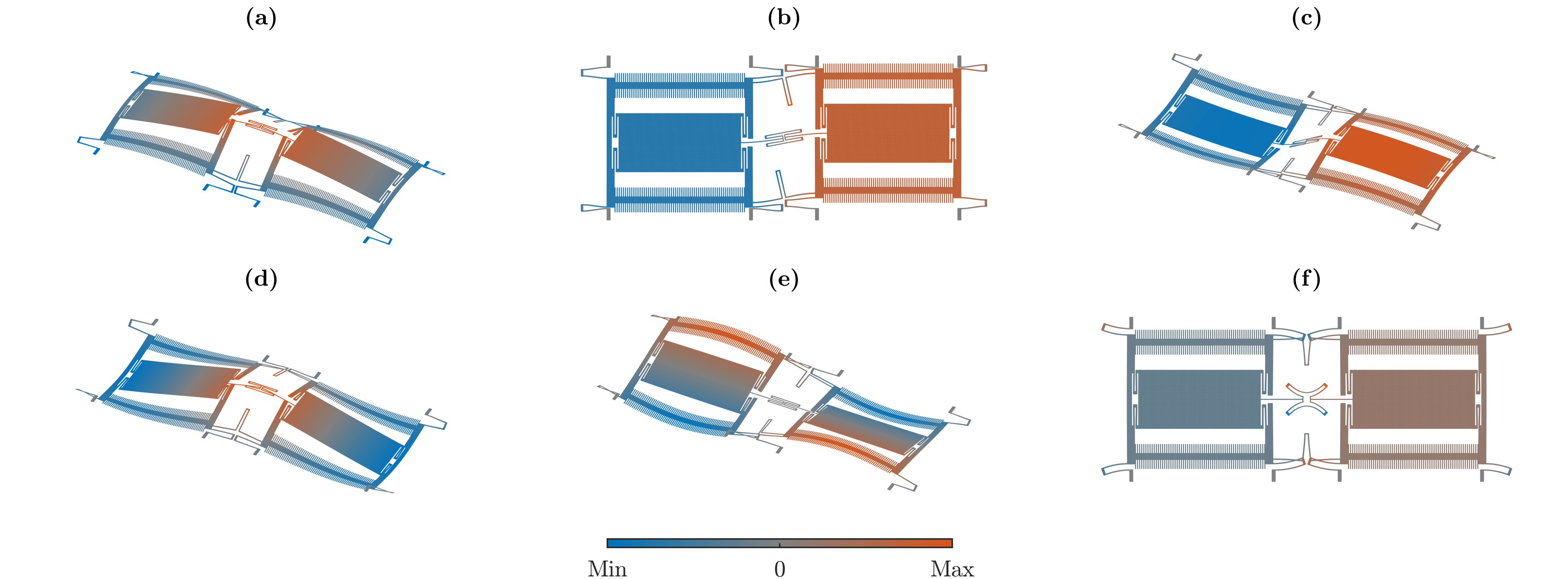}
	\caption{Mode shapes of the initial design. Only the modes relevant to the optimizations are displayed. For simplicity, only the displacement of the surface is shown. The coloring indicates the sum of the three displacement components at each point, normalized to each mode's maximum value. (a): Mode 1 with an eigenfrequency of \SI{22.0}{\kilo\hertz}. (b): Mode 3 (drive) with an eigenfrequency of \SI{24.4}{\kilo\hertz}. (c): Mode 4 (sense) with an eigenfrequency of \SI{27.3}{\kilo\hertz}. (d): Mode 7 with an eigenfrequency of \SI{51.0}{\kilo\hertz}. (e): Mode 9 with an eigenfrequency of \SI{55.7}{\kilo\hertz}. (f): Mode 13 with an eigenfrequency of \SI{70.3}{\kilo\hertz}.}
	\label{fig:ModeShapes_initial}
\end{figure*}
The single-axis MEMS gyroscope, that we will shape-optimize here, has already been subject of previous publications, which dealt with the modeling of nonlinearities \cite{putnik_simulation_2016,putnik_incorporating_2017,putnik_static_2017,putnik_predicting_2018,putnik_simulation_2018} and the shape optimization of eigenfrequencies \cite{schiwietz_shape_2024}, where a more detailed description of the model can be found. Figure~\ref{fig:InitialDesign} depicts a top view of the initial design, as the three-dimensional (3D) model is simply extruded along the out-of-plane direction. We assume that the entire mechanical structure is composed of linear isotropic polycrystalline silicon. The design exhibits a quarter symmetry in the $xy$-plane. In Fig.~\ref{fig:InitialDesign}, the substrate anchors are colored in blue, the springs in green and the comb drive electrodes in orange. Several eigenmodes of the initial design, which will be relevant in this work, are shown in Fig.~\ref{fig:ModeShapes_initial}. The operation of the MEMS gyroscope is based on the drive mode, shown in Fig.~\ref{fig:ModeShapes_initial}~(b), and the sense mode, shown in Fig.~\ref{fig:ModeShapes_initial}~(c). The relevance of the remaining modes in Fig.~\ref{fig:ModeShapes_initial} will be explained later on. An alternating voltage applied to the comb electrodes forces the drive mode to oscillate. An angular rate around the $x$-axis then excites the sense mode via the Coriolis force. The oscillation of the sense mode can be measured capacitively via electrodes underneath the masses.

\subsection{Optimization Problems} \label{sec:optimization_problems}
We will demonstrate the shape optimization of geometrically nonlinear 3-wave coupling coefficients with two different optimization problems. Both optimizations will be performed on the MEMS gyroscope in Fig.~\ref{fig:InitialDesign}. We apply the node-based shape optimization to the springs of the gyroscope. The design parameters are defined such that they shift the exterior nodes of the springs along the surface normals in the $xy$-plane. We formulate the optimization problems such that the squared norm of the design parameters $\bm{p}\cdot\bm{p}$ is minimized. By that, the optimization converges towards the design which is most similar to the initial design. The actual goals of the optimization will be introduced as constraints. In both optimization problems, we will enforce that the eigenfrequencies of drive mode $f_d$ and sense mode $f_s$ remain fixed within a tolerance. Additionally, we will constrain the mode shape of the drive mode $\bm{\phi}_{d}$ to remain parallel to the $y$-axis. This is motivated by the fact that the initial layout of a gyroscope is usually already designed such that drive and sense mode have desired eigenfrequencies and mode shapes, i.e., basic sensor functionality is ensured. Furthermore, manufacturability constraints on the minimum width of structures and minimum distance between structures will be employed. These are customary for MEMS resonators, since typical etching processes can not release structures which are arbitrarily close to each other or too wide.

The aim of the first optimization problem is the reduction of two different geometrically nonlinear 3-wave couplings between drive mode and two spurious modes by factors of 1000. We found that 3-wave couplings between in-plane, out-of-plane and torsional modes are typically large in our design. Therefore, we will decrease the 3-wave coupling coefficients $\bar{\alpha}_{d,1,9}=\tilde{\alpha}_{d,1,9}+\tilde{\alpha}_{d,9,1}$ and $\bar{\alpha}_{d,7,9}=\tilde{\alpha}_{d,7,9}+\tilde{\alpha}_{d,9,7}$ between the drive mode $d=3$, a torsional mode 9, and two distinct out-of-plane modes 1 and 7, see Fig.~\ref{fig:ModeShapes_initial}~(a), (b), (d) and (e). Note that $\bar{\alpha}_{d,1,9}$ and $\bar{\alpha}_{d,7,9}$ contain all nonlinear 3-wave couplings proportional to $q_{d}q_1q_9$ and $q_{d}q_7q_9$ in the potential energy, Eq.~\eqref{eq:strainenergy_modal}. We optimize their absolute values, as these determine the magnitude of the nonlinear coupling effects. The first optimization problem is formulated as
\begin{equation}
\begin{aligned}
\min_{\bm{p}}       & \quad  \bm{p}\cdot\bm{p} \\
\text{subject to: } & \quad |\bar{\alpha}_{d,1,9}| \leq 0.001\cdot|\bar{\alpha}_{d,1,9}|_0 \\
& \quad |\bar{\alpha}_{d,7,9}| \leq 0.001\cdot|\bar{\alpha}_{d,7,9}|_0 \\
& \quad 0.99f_{d,0} \leq f_{d} \leq 1.01f_{d,0} \\
& \quad 0.99f_{s,0} \leq f_{s} \leq 1.01f_{s,0} \\
& \quad \frac{\left\lVert \bm{\phi}_{d,x} \right\rVert}{\left\lVert \bm{\phi}_{d,y} \right\rVert} \leq 1.1 \frac{\left\lVert \bm{\phi}_{d,x} \right\rVert}{\left\lVert \bm{\phi}_{d,y} \right\rVert}_0 \ \text{in} \ V_{el} \\
& \quad d_j\geq\SI{2}{\micro\meter} \\ 
& \quad w_j\geq\SI{1.5}{\micro\meter} \\ 
& \quad -9 \leq p_j \leq 9 \\
& \quad j=1,\ldots,n_p \label{eq:OptimizationProblem1}
\end{aligned}
\end{equation}
where the subscript $0$ indicates quantities in the initial design, $\bm{\phi}_{d,x}$ and $\bm{\phi}_{d,y}$ refer to the $x$- and $y$-components of the drive mode's eigenvector, $V_{el}$ is the domain of the comb drive electrodes which is highlighted in orange in Fig.~\ref{fig:InitialDesign}, $d_j$ is the distance between structures, $w_j$ is the width of structures and $n_p=3413$ is the number of design parameters. The initial values of the 3-wave coupling coefficients were $|\bar{\alpha}_{d,1,9}|_0=\SI{4.1e18}{\frac{1}{\sqrt{\kilogram}\cdot\metre\cdot\second^2}}$ and $|\bar{\alpha}_{d,7,9}|_0=\SI{5.0e18}{\frac{1}{\sqrt{\kilogram}\cdot\metre\cdot\second^2}}$. The minimization in Eq.~\eqref{eq:OptimizationProblem1} ensures that the design is close to the initial design; the constraints on $|\bar{\alpha}_{d,1,9}|$ and $|\bar{\alpha}_{d,7,9}|$ reduce the magnitude of the 3-wave couplings by a factor of 1000 with respect to their initial values; the constraints on $f_{d}$ and $f_{s}$ keep the drive and sense mode eigenfrequencies fixed within $\pm1\%$ of their initial values; the constraint on the ratio between $\bm{\phi}_{d,x}$ and $\bm{\phi}_{d,y}$ ensures that the drive mode shape remains parallel to the $y$-axis within the comb drive electrodes, so that the sensor can be properly actuated; the constraints on $d_j$ and $w_j$ enforce that structures are at least \SI{2}{\micro\meter} apart and \SI{1.5}{\micro\meter} wide, to ensure the manufacturability with customary MEMS fabrication processes. The limits on $p_j$ were chosen such that a converged solution could be obtained without being unnecessarily large. Details on the calculation of $d_j$ and $w_j$ can be found in \cite{schiwietz_shape_2024}. 

In contrast to the first optimization, the second optimization problem will demonstrate the enhancement of geometrically nonlinear 3-wave effects. We will enforce a 1:2 internal resonance and increase the corresponding coupling coefficient by a factor of 250. The factor of 250 is due to the fact that this was roughly the largest coupling increase that we could obtain with our methodology. We consider the 1:2 coupling between drive mode and mode 7. Their mode shapes are shown in Fig.~\ref{fig:ModeShapes_initial}~(b) and (d). The second optimization problem reads
\begin{equation}
\begin{aligned}
\min_{\bm{p}}       & \quad  \bm{p}\cdot\bm{p} \\
\text{subject to: } & \quad |\tilde{\alpha}_{d,d,7}| \geq 250\cdot|\tilde{\alpha}_{d,d,7}|_0 \\
& \quad |2f_{d}-f_7|\leq\SI{100}{\hertz} \\
& \quad |f_7-f_{13}|\geq\SI{2}{\kilo\hertz} \\
& \quad 0.99f_{d,0} \leq f_{d} \leq 1.01f_{d,0} \\
& \quad 0.99f_{s,0} \leq f_{s} \leq 1.01f_{s,0} \\
& \quad \frac{\left\lVert \bm{\phi}_{d,x} \right\rVert}{\left\lVert \bm{\phi}_{d,y} \right\rVert} \leq 1.1 \frac{\left\lVert \bm{\phi}_{d,x} \right\rVert}{\left\lVert \bm{\phi}_{d,y} \right\rVert}_0 \ \text{in} \ V_{el} \\
& \quad d_j\geq\SI{2}{\micro\meter} \\ 
& \quad w_j\geq\SI{1.5}{\micro\meter} \\ 
& \quad -9 \leq p_j \leq 9 \\
& \quad j=1,\ldots,n_p \label{eq:OptimizationProblem2}
\end{aligned}
\end{equation}
where the constraint on $|\tilde{\alpha}_{d,d,7}|$ leads to an increase of the 1:2 internal resonance coupling coefficient between drive mode and mode 7 by a factor of 250 with respect to the initial value $|\tilde{\alpha}_{d,d,7}|_0=\SI{2.2e15}{\frac{1}{\sqrt{\kilogram}\cdot\metre\cdot\second^2}}$; the constraint on $|2f_{d}-f_7|$ enforces that the frequencies of drive mode and mode 7 deviate no more than \SI{100}{\hertz} from the ideal 1:2 internal resonance condition; the constraint on $|f_7-f_{13}|$ ensures that the eigenfrequencies of mode 7 and 13 remain at least \SI{2}{\kilo\hertz} apart to avoid mode mixing and the remaining constraints and minimization are as in Eq.~\eqref{eq:OptimizationProblem1}. Without the constraint on $|f_7-f_{13}|$, we found that modes 7 and 13 would mix. Looking at Fig.~\ref{fig:ModeShapes_initial}~(d) and (f), this meant that mode 7, which is initially an out-of-plane mode, also exhibited significant in-plane motion within the springs, similar to mode 13, leading to a large coupling $|\tilde{\alpha}_{d,d,7}|$. However, this would be an undesirable solution, as such an increase, based on mode mixing, would likely be very sensitive to the specific values of $f_7$ and $f_{13}$ and therefore to fabrication tolerances. Note that $\tilde{\alpha}_{d,d,7}$ contains all coupling coefficients that appear in the potential energy, Eq.~\eqref{eq:strainenergy_modal}, proportional to $q_{d}^2q_7$.

The sensitivities of the objective function and constraints in Eqs.~\eqref{eq:OptimizationProblem1} and \eqref{eq:OptimizationProblem2}, with respect to the design parameters, are required for the gradient-based shape optimization. The sensitivity of the objective function is trivially obtained. For all constraints that depend on eigenfrequencies but not on eigenvectors we calculate the sensitivities of the eigenfrequencies as in \cite{schiwietz_shape_2024} and obtain the sensitivities of the constraints via the chain rule. The sensitivities of the manufacturability constraints are also calculated as in \cite{schiwietz_shape_2024}. For all constraints that depend on eigenvectors we apply the adjoint method, according to Eqs.~\eqref{eq:lagrangiangradient}-\eqref{eq:nelson2}. The partial derivatives, required for the adjoint method, are detailed in Appendix~\ref{sec:A_derivatives}. During the shape optimization, the initial order of the modes might change. In order to consistently refer each constraint to the correct mode, we track the modes as described in Appendix~\ref{sec:A_modetracking}. We exploit the quarter symmetry of the initial design during the optimization. By that, the optimized designs also exhibit a quarter symmetry. All simulations and calculations were performed in a self-written FEM code in MATLAB.

\section{Results}\label{sec:results}
\begin{figure}
	\centering
	\includegraphics[width=0.9\textwidth]{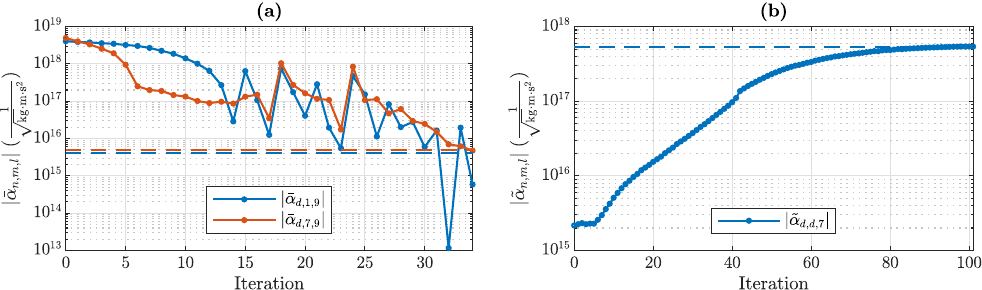}
	\caption{History of the optimized 3-wave coupling coefficients during the optimization process. The dashed lines indicate the target values. (a): First optimization problem, Eq.~\eqref{eq:OptimizationProblem1}. (b): Second optimization problem, Eq.~\eqref{eq:OptimizationProblem2}.}
	\label{fig:convergence}
\end{figure}
\begin{figure*}
	\centering
	\includegraphics[width=0.9\textwidth]{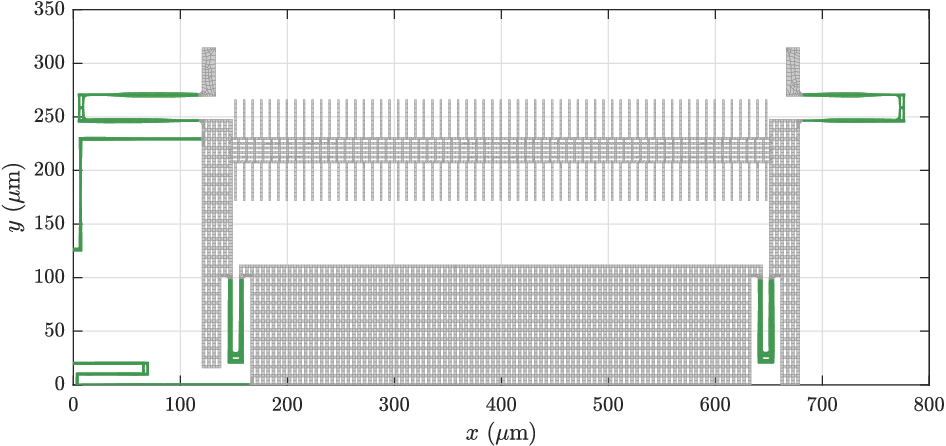}
	\caption{Final quarter model of the converged shape optimization according to Eq.~\eqref{eq:OptimizationProblem1}. The full design has a quarter symmetry, which is exploited in the optimization procedure. The springs are colored in green.}
	\label{fig:OptimizedQuarter_1}
\end{figure*}
\begin{figure*}
	\centering
	\includegraphics[width=0.9\textwidth]{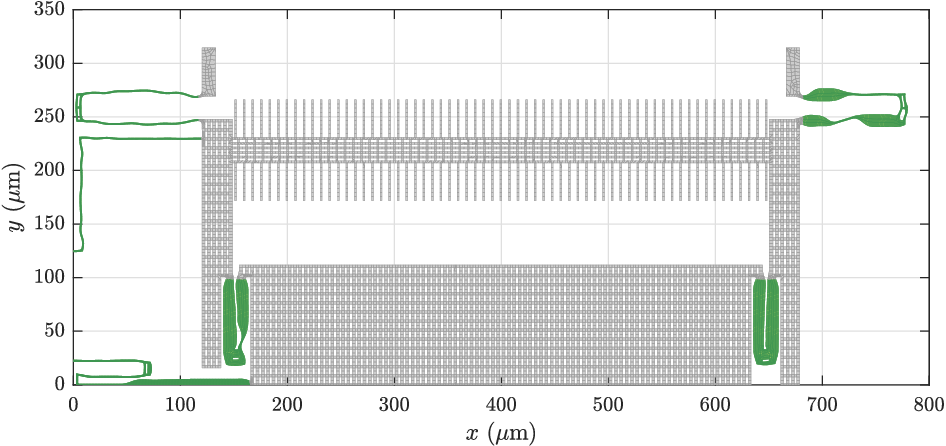}
	\caption{Final quarter model of the converged shape optimization according to Eq.~\eqref{eq:OptimizationProblem2}. The full design has a quarter symmetry, which is exploited in the optimization procedure. The springs are colored in green.}
	\label{fig:OptimizedQuarter_2}
\end{figure*}
\begin{figure}
	\centering
	\includegraphics[width=0.9\textwidth]{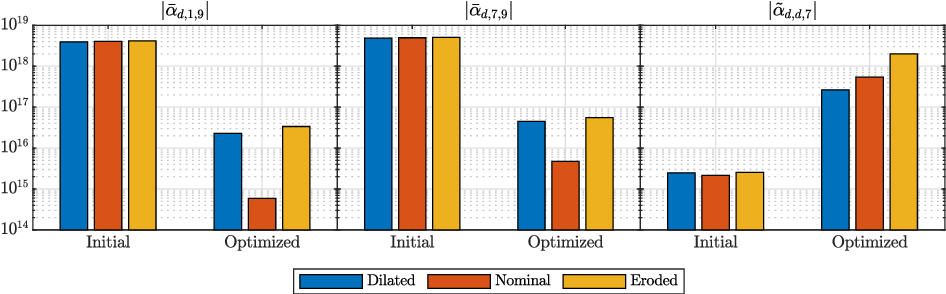}
	\caption{Dependence of the 3-wave coupling coefficients on process variations in the initial and optimized designs. Their values are shown for the nominal design, on which the optimization was based, as well as for the same design which was either dilated or eroded. The dilated design emulates an under-etching of the entire model such that all structures are thicker by \SI{100}{\nano\meter} in the xy-plane. The eroded design emulates an over-etching of the entire model such that all structures are thinner by \SI{100}{\nano\meter} in the xy-plane.}
	\label{fig:processdependence}
\end{figure}
The convergence history of the optimized 3-wave couplings for both optimization problems is shown in Fig.~\ref{fig:convergence}. The optimized quarter designs, corresponding to one quadrant of the complete structure depicted in Fig.~\ref{fig:InitialDesign}, are shown in Figs.~\ref{fig:OptimizedQuarter_1} and \ref{fig:OptimizedQuarter_2} for the first and second optimization problem, respectively. The mode shapes of the optimized designs, corresponding to the initial modes in Fig.~\ref{fig:ModeShapes_initial}, are shown in Appendix~\ref{sec:A_modeshapes}, for completeness. However, they do not deviate significantly from the initial mode shapes.

The first optimization problem, as stated in Eq.~\eqref{eq:OptimizationProblem1}, converged after 34 iterations. It can be seen in Fig.~\ref{fig:convergence}~(a) that the two depicted 3-wave couplings decreased continuously, except for a few iterations. Decreasing them by factors of 1000 took 34 iterations and the remaining constraints in Eq.~\eqref{eq:OptimizationProblem1} were already fulfilled in fewer iterations. We assume that the spikes in Fig.~\ref{fig:convergence}~(a) are due to the strong nonlinear dependence of the 3-wave couplings on the design parameters. A line search algorithm could be used to determine a step size which ensures reduction of the 3-wave coupling coefficients in every iteration. Looking at the optimized quarter layout in Fig.~\ref{fig:OptimizedQuarter_1}, we see rather subtle design changes. Nevertheless, the optimized 3-wave coupling coefficients were significantly reduced.

The second optimization problem, as stated in Eq.~\eqref{eq:OptimizationProblem2}, converged after 102 iterations. Again, most constraints were fulfilled in fewer iterations and the 3-wave coupling coefficient was the limiting factor in the optimization. Figure~\ref{fig:convergence}~(b) shows a rather smooth convergence of the 3-wave coupling coefficient towards the target value. The optimized quarter layout in Fig.~\ref{fig:OptimizedQuarter_2} exhibits significant morphing. Note that the shape changes are not just limited to variations in the width but also show significant bending of the initially straight springs. 

It appears that, in our examples, decreasing an initially large 3-wave coupling coefficient requires fewer iterations and smaller design changes than increasing an initially small 3-wave coupling coefficient. In both optimizations we find unintuitive design changes which have a significant impact on the optimized 3-wave coupling coefficients. This demonstrates the usefulness of gradient-based shape optimization for tuning geometrically nonlinear modal couplings without manual topology adjustments.

To investigate the sensitivity of the optimized designs to fabrication tolerances, we also calculate the coupling coefficients of the initial and optimized designs for a dilated and eroded geometry. This is a common approach when modelling MEMS or NEMS resonators, where the tolerances of the employed etching processes can result in variations of the in-plane width of structures \cite{sigmund2009manufacturing,hoj_ultra-coherent_2021}. For this purpose, we apply a uniform shift of \SI{+50}{\nano\meter} (dilated design) or \SI{-50}{\nano\meter} (eroded design) to all exterior nodes along their outward surface normal direction in the xy-plane. For example, a \SI{+50}{\nano\meter} shift will widen a spring by \SI{+100}{\nano\meter}, due to it being applied on both sides of the spring, which is a typical order of magnitude for fabrication tolerances in MEMS. The results for the minimized coupling coefficients, as shown in Fig.~\ref{fig:convergence}~(a), and the maximized coupling coefficient, as shown in Fig.~\ref{fig:convergence}~(b), are summarized in Fig.~\ref{fig:processdependence}. The results are shown for the initial and the optimized designs for a dilated (thicker), eroded (thinner) and nominal geometry. The initial nominal geometry corresponds to Fig.~\ref{fig:InitialDesign} and the optimized nominal geometries correspond to Figs.~\ref{fig:OptimizedQuarter_1} and \ref{fig:OptimizedQuarter_2}. One can see a clear impact of the fabrication tolerances on the optimized 3-wave coupling coefficients, whereas the effect in the initial design is rather small. The coupling coefficients that were reduced, as shown in Fig.~\ref{fig:convergence}~(a), increase by roughly one order of magnitude in the optimized design for the eroded and dilated geometry compared to the nominal design, as seen in Fig.~\ref{fig:processdependence}. Nevertheless, they still remain around two orders of magnitude below the initial values. The coupling coefficient that was increased in the second optimization, as seen in Fig.~\ref{fig:convergence}~(b), changes by less than one order of magnitude for the eroded and dilated geometry compared to the nominal design, as seen in Fig.~\ref{fig:processdependence}. Its value remains roughly two orders of magnitude larger than in the initial design.

Our results highlight that even with fabrication errors the nonlinear coupling coefficients can be tuned over various orders of magnitude. Note that we did not consider the fabrication errors during the optimization. A common approach is to evaluate the dilated, nominal and eroded design and to optimize the worst case out of the three designs in each iteration \cite{sigmund2009manufacturing,hoj_ultra-coherent_2021}. Fabricating and experimentally verifying such structures will be subject of future research. For that purpose, it would be advantageous to consider the fabrication errors during the optimization, to obtain a robust design.

\section{Conclusion}\label{sec:conclusion}
In this work, we developed a framework for the node-based shape optimization of geometrically nonlinear 3-wave coupling coefficients, and applied it to the design of a MEMS gyroscope. It was demonstrated that 3-wave coupling coefficients can be increased or decreased by several orders of magnitude with our approach, while also fulfilling manufacturability constraints and constraints on eigenfrequencies. The 3-wave coupling coefficients between in-plane, out-of-plane and torsional modes could be reduced by 3 orders of magnitude with very subtle design changes. The 3-wave coupling coefficient leading to an 1:2 internal resonance between an in-plane and an out-of-plane mode could be increased by a factor of 250, while also enforcing the corresponding resonance condition. The optimizations lead to design changes which are non-intuitive even to experienced human designers. Furthermore, it was estimated that, even in the presence of fabrication errors, the optimized coupling coefficients were still around two orders of magnitude smaller or larger than the initial values. Our results highlight that shape optimization is a powerful design tool for MEMS and NEMS resonator applications which exploit nonlinearities as well as applications where nonlinearities lead to parasitic effects. In the future, it would be intriguing to fabricate and characterize test structures with optimized geometrically nonlinear 3-wave couplings, to experimentally prove the applicability of our approach. For that purpose it would also be benficial to consider fabrication errors during optimization to obtain robust designs.

\section*{Data Availability}
All the data are available in the manuscript, or they will be provided upon reasonable request. 

\bibliography{main}

\begin{thebibliography}{10}
\urlstyle{rm}
\expandafter\ifx\csname url\endcsname\relax
  \def\url#1{\texttt{#1}}\fi
\expandafter\ifx\csname urlprefix\endcsname\relax\def\urlprefix{URL }\fi
\expandafter\ifx\csname doiprefix\endcsname\relax\def\doiprefix{DOI: }\fi
\providecommand{\bibinfo}[2]{#2}
\providecommand{\eprint}[2][]{\url{#2}}

\bibitem{bachtold_mesoscopic_2022}
\bibinfo{author}{Bachtold, A.}, \bibinfo{author}{Moser, J.} \&
  \bibinfo{author}{Dykman, M.}
\newblock \bibinfo{journal}{\bibinfo{title}{Mesoscopic physics of
  nanomechanical systems}}.
\newblock {\emph{\JournalTitle{Reviews of Modern Physics}}}
  \textbf{\bibinfo{volume}{94}}, \bibinfo{pages}{045005},
  \doiprefix\url{10.1103/RevModPhys.94.045005} (\bibinfo{year}{2022}).

\bibitem{eichler_classical_2023}
\bibinfo{author}{Eichler, A.} \& \bibinfo{author}{Zilberberg, O.}
\newblock \emph{\bibinfo{title}{Classical and quantum parametric phenomena}}
  (\bibinfo{publisher}{Oxford University Press}, \bibinfo{address}{New York},
  \bibinfo{year}{2023}).

\bibitem{bukhari2021novel}
\bibinfo{author}{Bukhari, S. A.~R.}, \bibinfo{author}{Saleem, M.~M.},
  \bibinfo{author}{Hamza, A.} \& \bibinfo{author}{Bazaz, S.~A.}
\newblock \bibinfo{journal}{\bibinfo{title}{A novel design of high resolution
  mems gyroscope using mode-localization in weakly coupled resonators}}.
\newblock {\emph{\JournalTitle{IEEE Access}}} \textbf{\bibinfo{volume}{9}},
  \bibinfo{pages}{157597--157608} (\bibinfo{year}{2021}).

\bibitem{miao2022nonlinearity}
\bibinfo{author}{Miao, T.} \emph{et~al.}
\newblock \bibinfo{journal}{\bibinfo{title}{Nonlinearity-mediated digitization
  and amplification in electromechanical phonon-cavity systems}}.
\newblock {\emph{\JournalTitle{Nature Communications}}}
  \textbf{\bibinfo{volume}{13}}, \bibinfo{pages}{2352} (\bibinfo{year}{2022}).

\bibitem{nayfeh2008nonlinear}
\bibinfo{author}{Nayfeh, A.~H.} \& \bibinfo{author}{Mook, D.~T.}
\newblock \emph{\bibinfo{title}{Nonlinear oscillations}}
  (\bibinfo{publisher}{John Wiley \& Sons}, \bibinfo{year}{2008}).

\bibitem{kerschen_modal_2014}
\bibinfo{editor}{Kerschen, G.} (ed.) \emph{\bibinfo{title}{Modal {Analysis} of
  {Nonlinear} {Mechanical} {Systems}}}, vol. \bibinfo{volume}{555}
  (\bibinfo{publisher}{Springer}, \bibinfo{address}{Vienna},
  \bibinfo{year}{2014}).

\bibitem{touze_nonlinear_2006}
\bibinfo{author}{Touz{\'e}, C.} \& \bibinfo{author}{Amabili, M.}
\newblock \bibinfo{journal}{\bibinfo{title}{Nonlinear normal modes for damped
  geometrically nonlinear systems: {Application} to reduced-order modelling of
  harmonically forced structures}}.
\newblock {\emph{\JournalTitle{Journal of Sound and Vibration}}}
  \textbf{\bibinfo{volume}{298}}, \bibinfo{pages}{958--981},
  \doiprefix\url{10.1016/j.jsv.2006.06.032} (\bibinfo{year}{2006}).

\bibitem{opreni_model_2021}
\bibinfo{author}{Opreni, A.}, \bibinfo{author}{Vizzaccaro, A.},
  \bibinfo{author}{Frangi, A.} \& \bibinfo{author}{Touz{\'e}, C.}
\newblock \bibinfo{journal}{\bibinfo{title}{Model order reduction based on
  direct normal form: application to large finite element {MEMS} structures
  featuring internal resonance}}.
\newblock {\emph{\JournalTitle{Nonlinear Dynamics}}}
  \textbf{\bibinfo{volume}{105}}, \bibinfo{pages}{1237--1272},
  \doiprefix\url{10.1007/s11071-021-06641-7} (\bibinfo{year}{2021}).

\bibitem{sarrafan_nonlinear_2019}
\bibinfo{author}{Sarrafan, A.}, \bibinfo{author}{Azimi, S.},
  \bibinfo{author}{Golnaraghi, F.} \& \bibinfo{author}{Bahreyni, B.}
\newblock \bibinfo{journal}{\bibinfo{title}{A {Nonlinear} {Rate} {Microsensor}
  utilising {Internal} {Resonance}}}.
\newblock {\emph{\JournalTitle{Scientific Reports}}}
  \textbf{\bibinfo{volume}{9}}, \bibinfo{pages}{8648},
  \doiprefix\url{10.1038/s41598-019-44669-3} (\bibinfo{year}{2019}).

\bibitem{nitzan_self-induced_2015}
\bibinfo{author}{Nitzan, S.~H.} \emph{et~al.}
\newblock \bibinfo{journal}{\bibinfo{title}{Self-induced parametric
  amplification arising from nonlinear elastic coupling in a micromechanical
  resonating disk gyroscope}}.
\newblock {\emph{\JournalTitle{Scientific Reports}}}
  \textbf{\bibinfo{volume}{5}}, \bibinfo{pages}{9036},
  \doiprefix\url{10.1038/srep09036} (\bibinfo{year}{2015}).

\bibitem{antonio_frequency_2012}
\bibinfo{author}{Antonio, D.}, \bibinfo{author}{Zanette, D.~H.} \&
  \bibinfo{author}{Lopez, D.}
\newblock \bibinfo{journal}{\bibinfo{title}{Frequency stabilization in
  nonlinear micromechanical oscillators}}.
\newblock {\emph{\JournalTitle{Nature Communications}}}
  \textbf{\bibinfo{volume}{3}}, \bibinfo{pages}{806},
  \doiprefix\url{10.1038/ncomms1813} (\bibinfo{year}{2012}).

\bibitem{nabholz_parametric_2020}
\bibinfo{author}{Nabholz, U.}, \bibinfo{author}{Lamprecht, L.},
  \bibinfo{author}{Mehner, J.~E.}, \bibinfo{author}{Zimmermann, A.} \&
  \bibinfo{author}{Degenfeld-Schonburg, P.}
\newblock \bibinfo{journal}{\bibinfo{title}{Parametric amplification of
  broadband vibrational energy harvesters for energy-autonomous sensors enabled
  by field-induced striction}}.
\newblock {\emph{\JournalTitle{Mechanical Systems and Signal Processing}}}
  \textbf{\bibinfo{volume}{139}}, \bibinfo{pages}{106642},
  \doiprefix\url{10.1016/j.ymssp.2020.106642} (\bibinfo{year}{2020}).

\bibitem{ganesan_phononic_2017}
\bibinfo{author}{Ganesan, A.}, \bibinfo{author}{Do, C.} \&
  \bibinfo{author}{Seshia, A.}
\newblock \bibinfo{journal}{\bibinfo{title}{Phononic {Frequency} {Comb} via
  {Intrinsic} {Three}-{Wave} {Mixing}}}.
\newblock {\emph{\JournalTitle{Physical Review Letters}}}
  \textbf{\bibinfo{volume}{118}}, \bibinfo{pages}{033903},
  \doiprefix\url{10.1103/PhysRevLett.118.033903} (\bibinfo{year}{2017}).

\bibitem{nabholz_nonlinear_2019}
\bibinfo{author}{Nabholz, U.}, \bibinfo{author}{Curcic, M.},
  \bibinfo{author}{Mehner, J.~E.} \& \bibinfo{author}{Degenfeld-Schonburg, P.}
\newblock \bibinfo{title}{Nonlinear {Dynamical} {System} {Model} for {Drive}
  {Mode} {Amplitude} {Instabilities} in {MEMS} {Gyroscopes}},
  \doiprefix\url{10.1109/ISISS.2019.8739703} (\bibinfo{year}{2019}).

\bibitem{nabholz_spontaneous_2019}
\bibinfo{author}{Nabholz, U.}, \bibinfo{author}{Schatz, F.},
  \bibinfo{author}{Mehner, J.~E.} \& \bibinfo{author}{Degenfeld-Schonburg, P.}
\newblock \bibinfo{journal}{\bibinfo{title}{Spontaneous {Parametric}
  {Down}-{Conversion} {Induced} by {Non}-{Degenerate} {Three}-{Wave} {Mixing}
  in a {Scanning} {MEMS} {Micro} {Mirror}}}.
\newblock {\emph{\JournalTitle{Scientific Reports}}}
  \textbf{\bibinfo{volume}{9}}, \bibinfo{pages}{3997},
  \doiprefix\url{10.1038/s41598-019-40377-0} (\bibinfo{year}{2019}).

\bibitem{nabholz_validating_2020}
\bibinfo{author}{Nabholz, U.}, \bibinfo{author}{Stockmar, F.},
  \bibinfo{author}{Mehner, J.~E.} \& \bibinfo{author}{Degenfeld-Schonburg, P.}
\newblock \bibinfo{journal}{\bibinfo{title}{Validating the {Critical} {Point}
  of {Spontaneous} {Parametric} {Downconversion} for {Over} 600 {Scanning}
  {MEMS} {Micromirrors} on {Wafer} {Level}}}.
\newblock {\emph{\JournalTitle{IEEE Sensors Letters}}}
  \textbf{\bibinfo{volume}{4}}, \bibinfo{pages}{1--4},
  \doiprefix\url{10.1109/LSENS.2020.2964384} (\bibinfo{year}{2020}).

\bibitem{lifshitz2008nonlinear}
\bibinfo{author}{Lifshitz, R.} \& \bibinfo{author}{Cross, M.~C.}
\newblock \bibinfo{journal}{\bibinfo{title}{Nonlinear dynamics of
  nanomechanical and micromechanical resonators}}.
\newblock {\emph{\JournalTitle{Reviews of nonlinear dynamics and complexity}}}
  \textbf{\bibinfo{volume}{1}}, \doiprefix\url{10.1002/9783527626359.ch1}
  (\bibinfo{year}{2008}).

\bibitem{giannini_size_2020}
\bibinfo{author}{Giannini, D.}, \bibinfo{author}{Bonaccorsi, G.} \&
  \bibinfo{author}{Braghin, F.}
\newblock \bibinfo{journal}{\bibinfo{title}{Size optimization of {MEMS}
  gyroscopes using substructuring}}.
\newblock {\emph{\JournalTitle{European Journal of Mechanics - A/Solids}}}
  \textbf{\bibinfo{volume}{84}}, \bibinfo{pages}{104045},
  \doiprefix\url{10.1016/j.euromechsol.2020.104045} (\bibinfo{year}{2020}).

\bibitem{giannini_topology_2020}
\bibinfo{author}{Giannini, D.}, \bibinfo{author}{Braghin, F.} \&
  \bibinfo{author}{Aage, N.}
\newblock \bibinfo{journal}{\bibinfo{title}{Topology optimization of {2D}
  in-plane single mass {MEMS} gyroscopes}}.
\newblock {\emph{\JournalTitle{Structural and Multidisciplinary Optimization}}}
  \textbf{\bibinfo{volume}{62}}, \bibinfo{pages}{2069--2089},
  \doiprefix\url{10.1007/s00158-020-02595-3} (\bibinfo{year}{2020}).

\bibitem{giannini_topology_2022}
\bibinfo{author}{Giannini, D.}, \bibinfo{author}{Aage, N.} \&
  \bibinfo{author}{Braghin, F.}
\newblock \bibinfo{journal}{\bibinfo{title}{Topology optimization of {MEMS}
  resonators with target eigenfrequencies and modes}}.
\newblock {\emph{\JournalTitle{European Journal of Mechanics - A/Solids}}}
  \textbf{\bibinfo{volume}{91}}, \bibinfo{pages}{104352},
  \doiprefix\url{10.1016/j.euromechsol.2021.104352} (\bibinfo{year}{2022}).

\bibitem{shin_spiderweb_2022}
\bibinfo{author}{Shin, D.} \emph{et~al.}
\newblock \bibinfo{journal}{\bibinfo{title}{Spiderweb {Nanomechanical}
  {Resonators} via {Bayesian} {Optimization}: {Inspired} by {Nature} and
  {Guided} by {Machine} {Learning}}}.
\newblock {\emph{\JournalTitle{Advanced Materials}}}
  \textbf{\bibinfo{volume}{34}}, \bibinfo{pages}{2106248},
  \doiprefix\url{10.1002/adma.202106248} (\bibinfo{year}{2022}).

\bibitem{hoj_ultra-coherent_2021}
\bibinfo{author}{Hoj, D.} \emph{et~al.}
\newblock \bibinfo{journal}{\bibinfo{title}{Ultra-coherent nanomechanical
  resonators based on inverse design}}.
\newblock {\emph{\JournalTitle{Nature Communications}}}
  \textbf{\bibinfo{volume}{12}}, \bibinfo{pages}{5766},
  \doiprefix\url{10.1038/s41467-021-26102-4} (\bibinfo{year}{2021}).

\bibitem{pedersen_designing_2005}
\bibinfo{author}{Pedersen, N.}
\newblock \bibinfo{journal}{\bibinfo{title}{Designing plates for minimum
  internal resonances}}.
\newblock {\emph{\JournalTitle{Structural and Multidisciplinary Optimization}}}
  \textbf{\bibinfo{volume}{30}}, \bibinfo{pages}{297--307},
  \doiprefix\url{10.1007/s00158-005-0529-x} (\bibinfo{year}{2005}).

\bibitem{schiwietz_shape_2024}
\bibinfo{author}{Schiwietz, D.}, \bibinfo{author}{H{\"o}rsting, M.},
  \bibinfo{author}{Weig, E.~M.}, \bibinfo{author}{Degenfeld-Schonburg, P.} \&
  \bibinfo{author}{Wenzel, M.}
\newblock \bibinfo{journal}{\bibinfo{title}{Shape {Optimization} of
  {Eigenfrequencies} in {MEMS} {Gyroscopes}}}.
\newblock {\emph{\JournalTitle{arXiv preprint arXiv:2402.05837}}}
  \doiprefix\url{10.48550/ARXIV.2402.05837} (\bibinfo{year}{2024}).

\bibitem{dou_structural_2015}
\bibinfo{author}{Dou, S.}, \bibinfo{author}{Strachan, B.~S.},
  \bibinfo{author}{Shaw, S.~W.} \& \bibinfo{author}{Jensen, J.~S.}
\newblock \bibinfo{journal}{\bibinfo{title}{Structural optimization for
  nonlinear dynamic response}}.
\newblock {\emph{\JournalTitle{Philosophical Transactions of the Royal Society
  A: Mathematical, Physical and Engineering Sciences}}}
  \textbf{\bibinfo{volume}{373}}, \bibinfo{pages}{20140408},
  \doiprefix\url{10.1098/rsta.2014.0408} (\bibinfo{year}{2015}).

\bibitem{li_tailoring_2017}
\bibinfo{author}{Li, L.~L.} \emph{et~al.}
\newblock \bibinfo{journal}{\bibinfo{title}{Tailoring the nonlinear response of
  {MEMS} resonators using shape optimization}}.
\newblock {\emph{\JournalTitle{Applied Physics Letters}}}
  \textbf{\bibinfo{volume}{110}}, \bibinfo{pages}{081902},
  \doiprefix\url{10.1063/1.4976749} (\bibinfo{year}{2017}).

\bibitem{zega2019hardening}
\bibinfo{author}{Zega, V.}, \bibinfo{author}{Langfelder, G.},
  \bibinfo{author}{Falorni, L.~G.} \& \bibinfo{author}{Comi, C.}
\newblock \bibinfo{journal}{\bibinfo{title}{Hardening, softening, and linear
  behavior of elastic beams in mems: An analytical approach}}.
\newblock {\emph{\JournalTitle{Journal of Microelectromechanical Systems}}}
  \textbf{\bibinfo{volume}{28}}, \bibinfo{pages}{189--198}
  (\bibinfo{year}{2019}).

\bibitem{wriggers_nonlinear_2008}
\bibinfo{author}{Wriggers, P.}
\newblock \emph{\bibinfo{title}{Nonlinear {Finite} {Element} {Methods}}}
  (\bibinfo{publisher}{Springer Berlin Heidelberg}, \bibinfo{address}{Berlin,
  Heidelberg}, \bibinfo{year}{2008}).

\bibitem{touze_model_2021}
\bibinfo{author}{Touze, C.}, \bibinfo{author}{Vizzaccaro, A.} \&
  \bibinfo{author}{Thomas, O.}
\newblock \bibinfo{journal}{\bibinfo{title}{Model order reduction methods for
  geometrically nonlinear structures: a review of nonlinear techniques}}.
\newblock {\emph{\JournalTitle{Nonlinear Dynamics}}}
  \textbf{\bibinfo{volume}{105}}, \bibinfo{pages}{1141--1190},
  \doiprefix\url{10.1007/s11071-021-06693-9} (\bibinfo{year}{2021}).

\bibitem{svanberg_method_1987}
\bibinfo{author}{Svanberg, K.}
\newblock \bibinfo{journal}{\bibinfo{title}{The method of moving asymptotes-a
  new method for structural optimization}}.
\newblock {\emph{\JournalTitle{International Journal for Numerical Methods in
  Engineering}}} \textbf{\bibinfo{volume}{24}}, \bibinfo{pages}{359--373},
  \doiprefix\url{10.1002/nme.1620240207} (\bibinfo{year}{1987}).

\bibitem{lee_adjoint_1999}
\bibinfo{author}{Lee, T.~H.}
\newblock \bibinfo{journal}{\bibinfo{title}{An adjoint variable method for
  structural design sensitivity analysis of a distinct eigenvalue problem}}.
\newblock {\emph{\JournalTitle{KSME International Journal}}}
  \textbf{\bibinfo{volume}{13}}, \bibinfo{pages}{470--476},
  \doiprefix\url{10.1007/BF02947716} (\bibinfo{year}{1999}).

\bibitem{tcherniak_topology_2002}
\bibinfo{author}{Tcherniak, D.}
\newblock \bibinfo{journal}{\bibinfo{title}{Topology optimization of resonating
  structures using {SIMP} method}}.
\newblock {\emph{\JournalTitle{International Journal for Numerical Methods in
  Engineering}}} \textbf{\bibinfo{volume}{54}}, \bibinfo{pages}{1605--1622},
  \doiprefix\url{10.1002/nme.484} (\bibinfo{year}{2002}).

\bibitem{nelson_simplified_1976}
\bibinfo{author}{Nelson, R.~B.}
\newblock \bibinfo{journal}{\bibinfo{title}{Simplified calculation of
  eigenvector derivatives}}.
\newblock {\emph{\JournalTitle{AIAA Journal}}} \textbf{\bibinfo{volume}{14}},
  \bibinfo{pages}{1201--1205}, \doiprefix\url{10.2514/3.7211}
  (\bibinfo{year}{1976}).

\bibitem{putnik_simulation_2016}
\bibinfo{author}{Putnik, M.}, \bibinfo{author}{Cardanobile, S.},
  \bibinfo{author}{Nagel, C.}, \bibinfo{author}{Degenfeld-Schonburg, P.} \&
  \bibinfo{author}{Mehner, J.}
\newblock \bibinfo{journal}{\bibinfo{title}{Simulation and {Modelling} of the
  {Drive} {Mode} {Nonlinearity} in {MEMS}-gyroscopes}}.
\newblock {\emph{\JournalTitle{Procedia Engineering}}}
  \textbf{\bibinfo{volume}{168}}, \bibinfo{pages}{950--953},
  \doiprefix\url{10.1016/j.proeng.2016.11.313} (\bibinfo{year}{2016}).

\bibitem{putnik_incorporating_2017}
\bibinfo{author}{Putnik, M.} \emph{et~al.}
\newblock \bibinfo{title}{Incorporating geometrical nonlinearities in reduced
  order models for {MEMS} gyroscopes},
  \doiprefix\url{10.1109/ISISS.2017.7935656} (\bibinfo{year}{2017}).

\bibitem{putnik_static_2017}
\bibinfo{author}{Putnik, M.} \emph{et~al.}
\newblock \bibinfo{title}{A static approach for the frequency shift of
  parasitic excitations in {MEMS} gyroscopes with geometric nonlinear drive
  mode}, \doiprefix\url{10.1109/ISISS.2017.7935651} (\bibinfo{year}{2017}).

\bibitem{putnik_predicting_2018}
\bibinfo{author}{Putnik, M.} \emph{et~al.}
\newblock \bibinfo{journal}{\bibinfo{title}{Predicting the {Resonance}
  {Frequencies} in {Geometric} {Nonlinear} {Actuated} {MEMS}}}.
\newblock {\emph{\JournalTitle{Journal of Microelectromechanical Systems}}}
  \textbf{\bibinfo{volume}{27}}, \bibinfo{pages}{954--962},
  \doiprefix\url{10.1109/JMEMS.2018.2871080} (\bibinfo{year}{2018}).

\bibitem{putnik_simulation_2018}
\bibinfo{author}{Putnik, M.} \emph{et~al.}
\newblock \bibinfo{title}{Simulation methods for generating reduced order
  models of {MEMS} sensors with geometric nonlinear drive motion},
  \doiprefix\url{10.1109/ISISS.2018.8358112} (\bibinfo{year}{2018}).

\bibitem{sigmund2009manufacturing}
\bibinfo{author}{Sigmund, O.}
\newblock \bibinfo{journal}{\bibinfo{title}{Manufacturing tolerant topology
  optimization}}.
\newblock {\emph{\JournalTitle{Acta Mechanica Sinica}}}
  \textbf{\bibinfo{volume}{25}}, \bibinfo{pages}{227--239}
  (\bibinfo{year}{2009}).

\end{thebibliography}

\section*{Acknowledgements}
The IPCEI ME/CT project is supported by the Federal Ministry for Economic Affairs and Climate Action on the basis of a decision by the German Parliament, by the Ministry for Economic Affairs, Labor and Tourism of Baden-W\"urttemberg based on a decision of the State Parliament of Baden-W\"urttemberg, the Free State of Saxony on the basis of the budget adopted by the Saxon State Parliament, the Bavarian State Ministry for Economic Affairs, Regional Development and Energy and financed by the European Union - NextGenerationEU.

The authors are thankful to Martin Putnik at Robert Bosch GmbH for providing the model of the initial design.

\section*{Author contributions statement}
D.S. and P.D. crafted the original idea. D.S. implemented the method into code. D.S., M.H. and M.W. developed the underlying shape optimization tool. D.S. performed the optimization. D.S. wrote the manuscript. P.D. and E.W. supervised the work. All authors reviewed the manuscript.

\section*{Additional information}

\textbf{Competing interests:} The authors declare no competing interests.

\begin{appendices}
	
	\section{Sensitivities} \label{sec:A_derivatives}
	
	\subsection{3-Wave Coupling Coefficients} \label{sec:A_derivatives_1}
	For the geometrically nonlinear 3-wave coupling coefficients, defined in Eq.~\eqref{eq:alpha}, we obtain the partial derivatives for the adjoint method as
	\begin{equation} \label{eq:dalpha_df}
	\frac{\partial \alpha_{n,m,l}}{\partial f_i}=0,
	\end{equation}
	\begin{equation}
	\frac{\partial\alpha_{n,m,l}}{\partial p_j}=\sum_e\sum_p\left[\left(\left(\frac{\partial \bm{B}_\varepsilon^e}{\partial p_j}\bm{\phi}^e_n\right)^T\bm{D}\bm{\eta}_{m,l}^e+\frac{1}{2}\bm{\varepsilon}^e_n\bm{D}\frac{\partial\bm{B}^e_{\eta}\left(\bm{\phi}^e_m\right)}{\partial p_j}\bm{\phi}^e_l\right)\det(\bm{J}^e)+\left(\bm{\varepsilon}^e_n\right)^T\bm{D}\bm{\eta}_{m,l}^e\frac{\partial \det(\bm{J}^e)}{\partial p_j}\right]W_p,
	\end{equation}
	\begin{equation} \label{eq:dalpha_dphi}
	\frac{\partial\alpha_{n,m,l}}{\partial \bm{\phi}_i}=\sum_e\left(\bm{L}^e\right)^T\int_{V^e}\left[\vphantom{\frac{1}{2}}\left(\bm{B}^e_{\varepsilon}\right)^T\bm{D}\bm{\eta}_{m,l}^e\delta_{n,i}+\frac{1}{2}\left(\bm{B}^e_{\eta}\left(\bm{\phi}^e_l\right)\right)^T\bm{D}\bm{\varepsilon}^e_n\delta_{m,i}+\frac{1}{2}\left(\bm{B}^e_{\eta}\left(\bm{\phi}^e_m\right)\right)^T\bm{D}\bm{\varepsilon}^e_n\delta_{l,i}\right]\mathrm{d}V,
	\end{equation}
	where $\bm{J}^e$ is the Jacobian matrix of element $e$, $p$ is the integration point index, $W_p$ is the integration point weight, $\bm{L}^e$ is a matrix which maps the element eigenvector to the global eigenvector as $\bm{\phi}_i^e=\bm{L}^e\bm{\phi}_i$ and $\delta_{i,j}$ is the Kronecker delta. To derive Eq.~\eqref{eq:dalpha_dphi}, we used the property $\bm{B}^e_{\eta}\left(\bm{\phi}^e_i\right)\bm{\phi}^e_j=\bm{B}^e_{\eta}\left(\bm{\phi}^e_j\right)\bm{\phi}^e_i$. The required sensitivities of $\bm{B}_\varepsilon^e$ and $\det(\bm{J}^e)$ are defined as in \cite{schiwietz_shape_2024}. Note that $\bm{B}^e_{\eta}\left(\bm{\phi}^e_i\right)=\bm{B}^e_{\eta}\left(\bm{H}_i^e,\bm{B}^e\right)$ is bilinear in $\bm{H}_i^e$ and $\bm{B}^e$, with $\bm{H}_i^e$ being the displacement gradient of the eigenvector of mode $i$ inside element $e$ and $\bm{B}^e$ containing shape function derivatives. The displacement gradient and its partial derivative can be obtained as
	\begin{align}
	\bm{H}_i^e&=\bm{\varPhi}_i^e\left(\bm{B}^e\right)^T, \\
	\frac{\partial \bm{H}_i^e}{\partial p_j}&=\bm{\varPhi}_i^e\left(\frac{\partial \bm{B}^e}{\partial p_j}\right)^T,
	\end{align}
	where $\bm{\varPhi}_i^e$ is a matrix in which each column corresponds to one node of element $e$ and contains the 3 eigenvector components belonging to that node for mode $i$. The matrices $\bm{B}^e$ and $\frac{\partial \bm{B}^e}{\partial p_j}$ are defined as in \cite{schiwietz_shape_2024}. Exploiting the bilinearity of $\bm{B}^e_{\eta}\left(\bm{\phi}^e_i\right)$, its partial derivative is obtained via the product rule as
	\begin{equation}
	\frac{\partial\bm{B}^e_{\eta}\left(\bm{\phi}^e_i\right)}{\partial p_j} = \bm{B}^e_{\eta}\left(\frac{\partial \bm{H}_i^e}{\partial p_j},\bm{B}^e\right) + \bm{B}^e_{\eta}\left(\bm{H}_i^e,\frac{\partial \bm{B}^e}{\partial p_j}\right).
	\end{equation}
	After evaluating Eqs.~\eqref{eq:dalpha_df}-\eqref{eq:dalpha_dphi} for all relevant $\alpha_{n,m,l}$, the corresponding partial derivatives of the in Section~\ref{sec:optimization_problems} introduced $|\bar{\alpha}_{d,1,9}|$, $|\bar{\alpha}_{d,7,9}|$ and $|\tilde{\alpha}_{d,d,7}|$ are obtained via the chain rule and their total sensitivities can be evaluated using the adjoint method.
	
	\subsection{Drive Mode Eigenvector Constraint} \label{sec:A_derivatives_2}
	The sensitivity of the constraint on the drive mode's eigenvector is required. Note that $\bm{\phi}_{d,x}$ and $\bm{\phi}_{d,y}$ have the same dimensions as $\bm{\phi}_{d}$ but simply contain zeros everywhere, except for the $x$ or $y$ degrees of freedom inside $V_{el}$, where they contain the corresponding values of $\bm{\phi}_{d}$. The sensitivity is calculated via the adjoint method and the required partial derivatives are
	\begin{align}
	\frac{\partial\left\lVert \bm{\phi}_{d,x} \right\rVert/\left\lVert \bm{\phi}_{d,y} \right\rVert}{\partial f_d} &= 0, \\
	\frac{\partial\left\lVert \bm{\phi}_{d,x} \right\rVert/\left\lVert \bm{\phi}_{d,y} \right\rVert}{\partial p_j} &= 0, \\
	\frac{\partial\left\lVert \bm{\phi}_{d,x} \right\rVert/\left\lVert \bm{\phi}_{d,y} \right\rVert}{\partial \bm{\phi}_{d}} &= \frac{\bm{\phi}_{d,x}}{\left\lVert \bm{\phi}_{d,x} \right\rVert\left\lVert \bm{\phi}_{d,y} \right\rVert} - \frac{\bm{\phi}_{d,y}\left\lVert \bm{\phi}_{d,x} \right\rVert}{\left\lVert \bm{\phi}_{d,y} \right\rVert^3}.
	\end{align}
	
	\section{Mode Tracking}\label{sec:A_modetracking}
	We employ a similar approach to mode tracking as in \cite{schiwietz_shape_2024}. We define a modal assurance criterion (MAC) as
	\begin{equation}
	\text{MAC}^{\{k\}}_{ij}= A\left(\bm{R}^{\{k\}}\bm{\phi}^{\{k\}}_i\right)^T\left(\bm{R}^{\{k-1\}}\bm{\phi}^{\{k-1\}}_j\right)+B\left(\bm{R}^{\{k\}}\bm{\phi}^{\{k\}}_i\right)^T\left(\bm{R}^{\{0\}}\bm{\phi}^{\{0\}}_j\right), 
	\label{eq:MAC}
	\end{equation}
	where $\bm{R}$ is the upper triangular matrix obtained from the Cholesky decomposition $\bm{M}=\bm{R}^T\bm{R}$, $A$ and $B$ are weights, the superscript $\{k\}$ refers to quantities calculated in the current iteration $k$, $\{k-1\}$ refers to the previous iteration and $\{0\}$ to the initial design. The mode $i$ for which $\text{MAC}^{\{k\}}_{ij}$ has the largest magnitude determines the mode number in iteration $k$ corresponding to mode $j$ from the previous iteration. We use $A=0.1$ and $B=0.9$, which we found to ensure a robust identification of the initially defined modes in this work, even if modes mix and demix throughout the optimization.
	
	\section{Mode Shapes} \label{sec:A_modeshapes}
\begin{figure*}
		\centering
		\includegraphics[width=\textwidth]{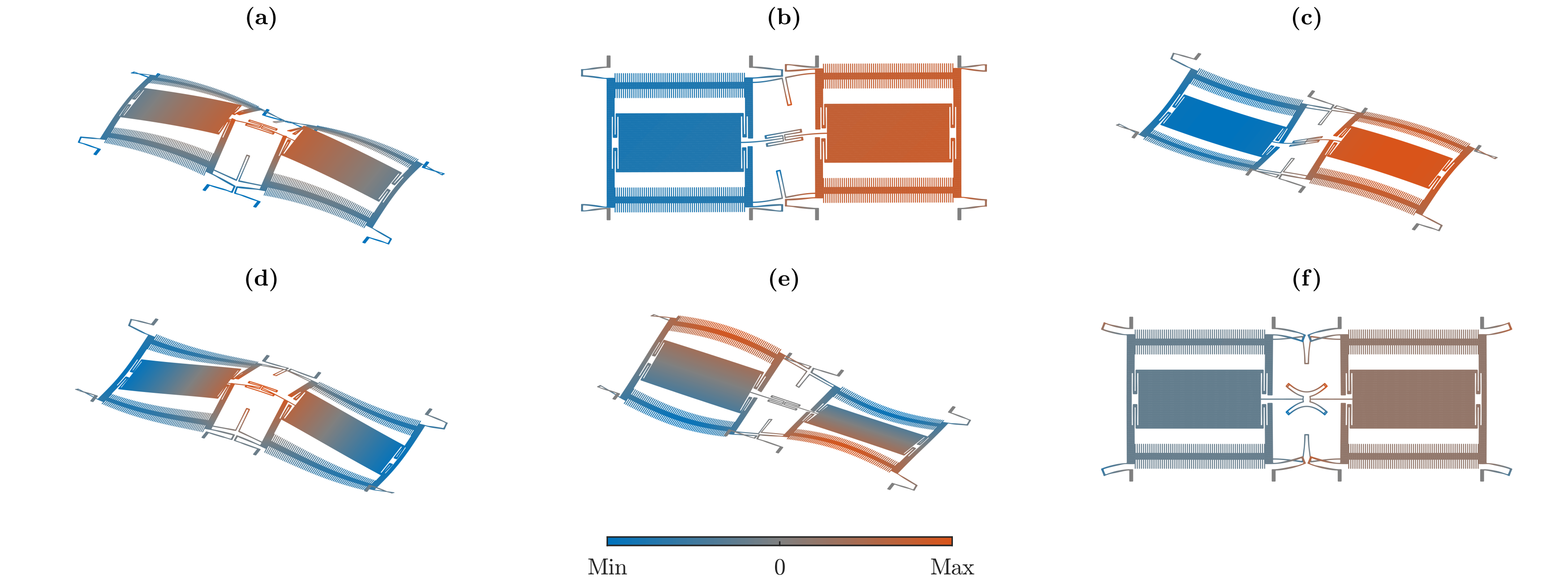}
		\caption{Mode shapes of the optimized design according to Eq.~\eqref{eq:OptimizationProblem1}. For simplicity, only the displacement of the surface is shown. The coloring indicates the sum of the three displacement components at each point, normalized to each mode's maximum value. We refer to the modes by their numbers in the initial design. (a): Mode 1 with an eigenfrequency of \SI{21.4}{\kilo\hertz}. (b): Mode 3 (drive) with an eigenfrequency of \SI{24.7}{\kilo\hertz}. (c): Mode 4 (sense) with an eigenfrequency of \SI{27.1}{\kilo\hertz}. (d): Mode 7 with an eigenfrequency of \SI{49.8}{\kilo\hertz}. (e): Mode 9 with an eigenfrequency of \SI{52.5}{\kilo\hertz}. (f): Mode 13 with an eigenfrequency of \SI{105.5}{\kilo\hertz}.}
		\label{fig:ModeShapes_opti_1}
	\end{figure*}

\begin{figure*}
		\centering
		\includegraphics[width=\textwidth]{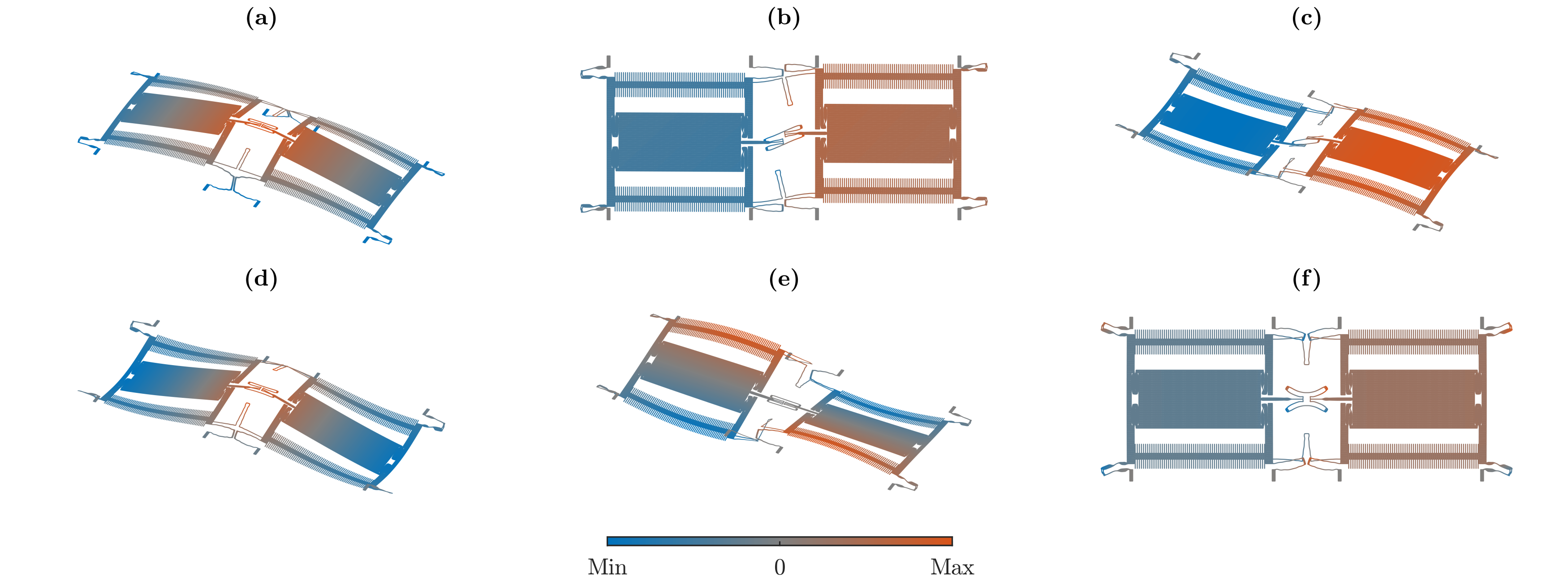}
		\caption{Mode shapes of the optimized design according to Eq.~\eqref{eq:OptimizationProblem2}. For simplicity, only the displacement of the surface is shown. The coloring indicates the sum of the three displacement components at each point, normalized to each mode's maximum value. We refer to the modes by their numbers in the initial design. (a): Mode 1 with an eigenfrequency of \SI{15.8}{\kilo\hertz}. (b): Mode 3 (drive) with an eigenfrequency of \SI{24.7}{\kilo\hertz}. (c): Mode 4 (sense) with an eigenfrequency of \SI{27.1}{\kilo\hertz}. (d): Mode 7 with an eigenfrequency of \SI{49.4}{\kilo\hertz}. (e): Mode 9 with an eigenfrequency of \SI{45.3}{\kilo\hertz}. (f): Mode 13 with an eigenfrequency of \SI{51.4}{\kilo\hertz}.}
		\label{fig:ModeShapes_opti_2}
	\end{figure*}

	Figures~\ref{fig:ModeShapes_opti_1} and \ref{fig:ModeShapes_opti_2} show the modes of the two optimized designs, corresponding to the modes in Fig.~\ref{fig:ModeShapes_initial}. Note that the order of the modes has changed in the optimized designs, when sorted by eigenfrequencies. However, for comparability, we label them in the same order as in the initial design.
	
\end{appendices}

\end{document}